\documentclass[11pt]{article}
\usepackage[margin=1in]{geometry}
\usepackage{authblk}
\usepackage{booktabs}
\usepackage{graphicx}
\usepackage{microtype}
\usepackage[hidelinks]{hyperref}
\usepackage{caption}

\hypersetup{
  pdftitle={A Symmetric Layer--Union Audit of Component Collapse in Hierarchical Procedural Corpora},
  pdfauthor={Jiuyi Zheng and Gan Xu}
}

\newcommand{\QualifyingSources}{2}
\newcommand{\EligibleSources}{6}
\newcommand{\QualifyingRatio}{2/6}
\newcommand{\QualifyingFamilies}{2}
\newcommand{\CandidateSourcesScreened}{5}
\newcommand{\AdditionalQualifyingSources}{1}
\newcommand{\StressCriterion}{A1}
\newcommand{\CriteriaATwoToAFour}{A2--A4}
\newcommand{\DocDialContentShare}{1.31\%}
\newcommand{\DocDialContainerShare}{1.56\%}
\newcommand{\DocDialUnionShare}{98.80\%}
\newcommand{\DocDialUnionComponents}{18}
\newcommand{\DocDialUnionEffective}{1.024}
\newcommand{\MyFixitContentShare}{0.17\%}
\newcommand{\MyFixitContainerShare}{0.06\%}
\newcommand{\MyFixitUnionShare}{41.64\%}
\newcommand{\MyFixitUnionComponents}{5,375}
\newcommand{\MyFixitUnionEffective}{5.760}
\newcommand{\WIQAContentShare}{0.20\%}
\newcommand{\WIQAContainerShare}{0.39\%}
\newcommand{\WIQAUnionShare}{1.49\%}
\newcommand{\WIQAContentEdges}{63}

\newcommand{\AllAuditCriteria}{A1--A4}
\newcommand{\CriterionATwo}{A2}
\newcommand{\CriterionAThree}{A3}
\newcommand{\CriterionAFour}{A4}
\newcommand{\RegisteredCriterionCount}{4}
\newcommand{\SingleLayerPassRequirement}{3/4}
\newcommand{\UnionFailRequirement}{3/4}
\newcommand{\ThresholdAOne}{0.01}
\newcommand{\ThresholdATwo}{0.20}
\newcommand{\ThresholdAThree}{30}
\newcommand{\RegisteredFoldCount}{3}
\newcommand{\AThreeReserveMultiplier}{10}
\newcommand{\ThresholdAFourFraction}{1/3}
\newcommand{\AllUnionAOneFailures}{6}
\newcommand{\AdditionalUnionFailuresNeeded}{2}
\newcommand{\NonqualifyingSources}{4}
\newcommand{\QualifierATwoToAFourPattern}{3/3}
\newcommand{\NonqualifierATwoToAFourPattern}{0/3}
\newcommand{\IntermediatePatternCount}{0}
\newcommand{\NonzeroCutoffCount}{3}
\newcommand{\CutoffOneOfThree}{1/3}
\newcommand{\CutoffTwoOfThree}{2/3}
\newcommand{\CutoffThreeOfThree}{3/3}
\newcommand{\SingleCriterionCheckCount}{3}

\newcommand{\MechanismPanelSources}{6}
\newcommand{\MechanismQualifiers}{2}
\newcommand{\PredictorCrossRho}{0.943}
\newcommand{\BridgeOutcomeRho}{0.943}
\newcommand{\NaiveOutcomeRho}{0.829}
\newcommand{\RegisteredOutcomeRhoThreshold}{0.8}

\newcommand{\PredictorDiscordantPairs}{1}
\newcommand{\BridgeSeparationThreshold}{63.7698}
\newcommand{\BridgeSeparationSlack}{42.8783}
\newcommand{\BridgeQualifierMinimum}{106.648}
\newcommand{\BridgeNonqualifierMaximum}{20.891}
\newcommand{\NaiveSeparationThreshold}{28.4043}
\newcommand{\NaiveSeparationSlack}{10.8687}
\newcommand{\DeletionAuditRows}{6}

\newcommand{\TauSweepSources}{2}
\newcommand{\TauGridPoints}{8}
\newcommand{\TauGridLowerBound}{0.50}
\newcommand{\TauRawInvariantMetricCount}{0}
\newcommand{\TauRawAuditedMetricCount}{3}
\newcommand{\TauWinnerRows}{32}
\newcommand{\TauRawFlips}{2}
\newcommand{\TauHolmSurvivingFlips}{0}
\newcommand{\MyFixitTauTopOneRatio}{4.03}
\newcommand{\MyFixitEdgeTopOneRatio}{239.83}
\newcommand{\MyFixitTopOneRatioQuotient}{59.48}

\newcommand{\CoverageExposureCorpora}{2}
\newcommand{\MyFixitCoverageUnits}{235,549}

\newcommand{\MyFixitCoverageIncompleteUnits}{198,544}

\newcommand{\MyFixitCoverageUnresolvedPairs}{12,309,728}
\newcommand{\MyFixitCoverageNaivePairs}{14,582,229}
\newcommand{\MyFixitCoverageExposure}{0.844160}
\newcommand{\MyFixitCoverageIncompleteFraction}{0.842899}
\newcommand{\MyFixitCoverageFractionGap}{0.001261}
\newcommand{\MyFixitCoveragePercentagePointGap}{0.126}
\newcommand{\MyFixitCoverageCompletePairsPerUnit}{61.4}
\newcommand{\MyFixitCoverageIncompletePairsPerUnit}{62.0}
\newcommand{\XWLPCoverageUnits}{3,915}

\newcommand{\XWLPCoverageIncompleteUnits}{2,285}

\newcommand{\XWLPCoverageUnresolvedPairs}{27,420}
\newcommand{\XWLPCoverageNaivePairs}{45,350}
\newcommand{\XWLPCoverageExposure}{0.604631}
\newcommand{\XWLPCoverageIncompleteFraction}{0.583653}
\newcommand{\XWLPCoverageFractionGap}{0.020978}
\newcommand{\XWLPCoveragePercentagePointGap}{2.098}
\newcommand{\XWLPCoverageCompletePairsPerUnit}{11.0}
\newcommand{\XWLPCoverageIncompletePairsPerUnit}{12.0}

\newcommand{\LexicalAuditCorpora}{2}
\newcommand{\MyFixitUnresolvedUnits}{198,544}
\newcommand{\MyFixitStepCueCount}{75,247}
\newcommand{\MyFixitStepCueRate}{37.90\%}
\newcommand{\MyFixitContextCueCount}{195,027}
\newcommand{\MyFixitContextCueRate}{98.23\%}
\newcommand{\MyFixitToolsFieldCueCount}{88,757}
\newcommand{\MyFixitToolsFieldCueRate}{44.70\%}
\newcommand{\XWLPUnresolvedOperations}{2,285}
\newcommand{\XWLPSentenceCueCount}{1,137}
\newcommand{\XWLPSentenceCueRate}{49.76\%}
\newcommand{\RequestedUniverseIndex}{universe[0]}

\newcommand{\SectionNineComparatorRelationTypes}{3}
\newcommand{\SectionNineComparatorDatasets}{10}
\newcommand{\SectionNineComparatorSplitVariants}{3}
\newcommand{\SectionNineComparatorModels}{600}

\newcommand{\SectionNineDescriptivePanelFraction}{2/6}

\title{A Symmetric Layer--Union Audit of Component Collapse in Hierarchical
Procedural Corpora}
\author[1,*,$\dagger$]{Jiuyi Zheng}
\author[1,*]{Gan Xu}
\affil[1]{University of Missouri, Columbia, Missouri, USA\newline
\texttt{j.zheng@missouri.edu}\quad\texttt{gxzmd@missouri.edu}}
\date{}

\begin{document}
\maketitle
\begingroup
\renewcommand{\thefootnote}{\fnsymbol{footnote}}
\footnotetext[1]{These authors contributed equally.}
\footnotetext[2]{Corresponding author: \texttt{j.zheng@missouri.edu}.}
\endgroup

\begin{abstract}
Component-disjoint leakage control can group corpus units by content
similarity, hierarchical membership, or both. Guvenilir and Doğan previously
showed that merging relation types can create a giant component that obstructs
splitting; we do not claim this phenomenon as new. We examine it through a
symmetric audit of a content-near-duplicate layer, a common-container layer,
and their union in a fixed panel of six hierarchical procedural corpora.
MyFixit and Doc2Dial exhibit the individual-layer-pass/union-fail pattern
under the same operational criteria. The resulting two-of-six fraction
describes this deliberately constructed panel and is not a prevalence
estimate. A prespecified bridge-specific predictor is associated with the
pattern, but it is not distinguished from a registered union-density control;
the panel therefore does not identify a bridge-specific mechanism. Secondary
diagnostics bound the interpretation of threshold sensitivity, annotation
coverage, and lexical cues without extending those findings beyond their
recorded sources and definitions. The paper's contribution is a bounded
measurement and audit: it keeps relation families visible, evaluates their
individual and union component structures symmetrically, and reports negative
cases and mechanism limits. It proposes neither a new splitting algorithm nor
a general causal claim about relation unions.
\end{abstract}

\section{Introduction}
\label{sec:introduction}

Leakage control often begins by deciding which data items must not be split
across training and evaluation partitions. In a hierarchical procedural
corpus, multiple relations can justify grouping. Procedural steps may express
the same or near-duplicate content, and otherwise different steps may belong
to the same guide, article, document, paragraph, or protocol. Either relation
can be represented as a graph whose connected components are kept intact. If
the relations are merged before transitive closure, however, a short path that
alternates between relation types can join units that neither layer connects
on its own. The resulting quarantine groups may therefore have very different
capacity from those suggested by either layer separately.\footnote{Companion
aggregate artifact: \url{https://github.com/joy91269/hierarepair}.}

This interaction sits within a mature literature on similarity-aware and
leakage-reduced splitting. Lo-Hi, GraphPart, DataSAIL, PLINDER, and related
work construct graph- or cluster-based partitions and document the tension
between separation and data retention
\cite{steshin2023_p4r24,teufel2023_graphpart,joeres2025_p4r25,durairaj2024_plinder}.
The closest direct precedent identified in our bounded review is Guvenilir and
Doğan's drug--target-interaction study \cite{guvenilir2023_dti}. That work
combines multiple relation types, reports a giant component that prevents
component-disjoint splitting, and applies deletion and community-based
repair. We therefore do not claim the broad chain from merged relations to
split infeasibility as a new observation. Record-linkage work has likewise
long noted that an erroneous link can propagate through transitive closure
\cite{gu2003_p4r06}, and near-duplicate handling is an established concern in
information-retrieval evaluation \cite{froebe2020_nearduplicates}.

The narrower question studied here is a cross-domain measurement question.
When hierarchical procedural corpora provide both natural-language content
and an explicit document-like container, how do the component summaries of a
content-near-duplicate layer, a common-container layer, and their union
compare under a symmetric operational audit? The emphasis is on retaining the
identity of each edge family. A large union component is informative only
when it can be compared with each constituent configuration under the same
unit mapping, closure rule, summaries, and decision criteria.

We evaluate a fixed panel of \EligibleSources{} gate-eligible sources drawn
from several procedural domains and provenance families. For each source,
Section~\ref{sec:study-design} maps source records to procedural units,
hierarchical containers, and content fields. C2 links normalized exact and
near-duplicate content, C3 links membership in the same container, and C5 is
the transitive closure of their union. Component count, largest-component
share, and effective component count describe the resulting partitions. The
source-level rule asks whether both individual layers retain sufficient
component dispersion under criteria \AllAuditCriteria{} while their union
fails the same checks. MyFixit was already known to meet this pattern;
\CandidateSourcesScreened{} additional candidates were screened, so the panel
is outcome-enriched rather than a basis for population inference.

The measurement yields \QualifyingSources{} qualifiers among the
\EligibleSources{} eligible sources. MyFixit and Doc2Dial, drawn from
\QualifyingFamilies{} provenance families, show the individual-pass/union-fail
pattern. The other sources remain visible as negative cases. Accordingly,
\QualifyingRatio{} is reported only as a descriptive panel fraction, not as a
base rate or prevalence estimate.

The mechanism-discrimination criterion is not met. Although a
bridge-specific predictor tracks the qualifier pattern, the registered
union-density control satisfies the same separation and outcome-association
conditions. The observed configuration contrast therefore does not identify
a bridge-specific mechanism. The study retains this negative result rather
than using secondary diagnostics to rescue a stronger explanation.

The study contributes a symmetric layer-versus-union measurement, an explicit
mechanism-discrimination test, and tightly scoped secondary diagnostics. It
reports the source mapping and nonqualifying cases, states the operational
criteria in auditable component summaries, and preserves the narrow
threshold-grid, coverage-exposure, and lexical findings within their recorded
evidence domains. These contributions do not include a splitting algorithm, a
giant-component theorem, an entity-resolution method, or a causal account of
bridging.

The rest of the paper proceeds as follows. Section~\ref{sec:study-design}
defines the panel, typed layers, component summaries, and decision rules.
Sections~\ref{sec:main-result} and~\ref{sec:criteria} report the source-level
measurement and audit-rule checks. Sections~\ref{sec:mechanism-identifiability}
and~\ref{sec:tau-inertness} present the mechanism and threshold boundaries.
Sections~\ref{sec:coverage-exposure} and~\ref{sec:structured-lexical-silence}
report the diagnostic analyses. Sections~\ref{sec:related-work}--
\ref{sec:reproducibility} position and bound the evidence before
Section~\ref{sec:conclusion} concludes.

\section{Measurement design and operational definitions}
\label{sec:study-design}

\subsection{Panel and source-specific units}

The measurement treats each corpus as a collection of procedural units nested
inside source-defined containers. A unit is the smallest occurrence to which
the source supplies usable procedural text; a container is the document-like
object whose members should remain together under the structural relation.
The mapping is source-specific because the same abstract roles are represented
by different schemas. In Human Know-How, a unit is a label-bearing target of
a structured \texttt{has\_step} relation and its container is the parent
WikiHow article. In MyFixit, guide-step occurrences are nested in repair
guides. OpenPI process-step records are mapped to their WikiHow articles;
WIQA paragraph steps to ProPara procedure paragraphs; Doc2Dial non-heading
document spans to government-service documents; and X-WLP operation events to
wet-lab protocols. The corresponding label, step, span, or sentence field is
the natural-language content field used by the content relation.

The fixed mapping yields \EligibleSources{} gate-eligible sources: Human
Know-How, MyFixit, OpenPI, WIQA, Doc2Dial, and X-WLP. Eligibility requires both
a resolvable hierarchical container and a natural-language content field from
which the two relation families below can be defined. WHPG is excluded because
one malformed record prevents an article identifier from being resolved for
all of its units. COIN is retained only as an auxiliary diagnostic outside the
gate-eligible panel because its segment label is a closed-vocabulary taxonomy
item rather than natural-language procedural text. The panel is not a
probability sample. MyFixit entered as the previously known qualifying source,
and \CandidateSourcesScreened{} additional candidates were screened. Results
are interpreted as outcomes for this fixed panel, not as estimates of
population prevalence.

\paragraph{Panel composition for the six gate-eligible sources.}
The following unnumbered summary describes only panel size and container
composition; it does not display qualifier status or anticipate the outcome.
Means are computed from the full-precision unit and container counts. The
median column is included because all six frozen aggregate diagnostics contain
a non-null value.

\providecommand{\EditorialXWLPMeanUnitsPerContainer}{14.03}
\providecommand{\EditorialMyFixitMeanUnitsPerContainer}{12.40}
\providecommand{\EditorialHumanKnowHowMaximumUnitsPerContainer}{206}
\providecommand{\EditorialMyFixitMaximumUnitsPerContainer}{137}
\providecommand{\EditorialDocTwoDialMeanUnitsPerContainer}{64.76}
\providecommand{\EditorialDocTwoDialMaximumUnitsPerContainer}{494}
\providecommand{\EditorialForestQualifierLowerBound}{14.49}
\providecommand{\EditorialForestNonqualifierUpperBound}{6.96}
\providecommand{\EditorialForestSpearmanMinimum}{0.886}
\providecommand{\EditorialForestSpearmanMaximum}{0.943}

\begin{center}
\small
\begin{tabular}{lrrrrr}
\toprule
Source & Units & Containers & \multicolumn{3}{c}{Units per container} \\
\cmidrule(lr){4-6}
 & & & Mean & Median & Maximum \\
\midrule
Human Know-How & 80,286 & 9,967 & 8.06 & 6 & 206 \\
MyFixit & 235,549 & 18,995 & 12.40 & 9 & 137 \\
OpenPI & 4,050 & 810 & 5.00 & 5 & 6 \\
WIQA & 2,557 & 379 & 6.75 & 6 & 10 \\
Doc2Dial & 31,602 & 488 & 64.76 & 56 & 494 \\
X-WLP & 3,915 & 279 & 14.03 & 14 & 36 \\
\bottomrule
\end{tabular}
\end{center}

Eligibility is structural rather than a claim of genre homogeneity. Doc2Dial
provides task-oriented government-service documents, while the other panel
members include repair guides, instructional articles, procedure paragraphs,
and laboratory protocols. Doc2Dial satisfies the same formal requirements for
a unit, container, and content field. This subject-matter heterogeneity limits
external interpretation of the fixed panel but does not change source
eligibility.

\subsection{Typed relation layers and closure}

For each source, let $V$ be its mapped unit occurrences. We construct two
undirected edge families on the common vertex set $V$. The content layer (C2)
links units whose normalized content is identical or satisfies the registered
near-duplicate rule. Normalization case-folds Unicode text, replaces
punctuation and underscores with spaces, and collapses whitespace; the
non-exact comparison uses word-trigram sets and registered Jaccard similarity.
Very short strings connect only under normalized exact equality. The
container layer (C3) expands every source-defined container into a clique,
linking every pair of its units. Thus C2 represents reusable or repeated
content, whereas C3 represents membership in one article, guide, paragraph,
document, or protocol. The labels describe graph construction and do not
assert that every edge is a semantic equivalence or a causal dependency.

The union configuration C5 contains both C2 and C3 edges. In every
configuration, quarantine groups are the connected components obtained by
transitive closure: if any path connects two units, they belong to the same
indivisible group for a component-disjoint split. The comparison is symmetric.
C2, C3, and C5 are evaluated with the same component summaries and the same
operational rule. This design isolates an observed configuration contrast; it
does not treat the union as the only legitimate representation or infer a
general causal effect from that contrast.

\subsection{Component summaries and the source-level rule}

Suppose a configuration partitions $N$ units into components of sizes
$n_1,\ldots,n_m$, and write $p_j=n_j/N$. We report the component count $m$,
the largest-component share $\max_j p_j$, and the effective component count

\[
  N_{\mathrm{eff}}=\left(\sum_{j=1}^{m}p_j^2\right)^{-1}.
\]

Component count records how many indivisible groups remain,
largest-component share captures the largest immediate capacity constraint,
and effective component count summarizes concentration across the full
component-size distribution. These quantities are structural diagnostics;
the study does not construct or optimize a new train--test split.

The source-level rule applies criteria \AllAuditCriteria{} to C2, C3, and C5.
Criterion \StressCriterion{} requires largest-component share at most
\ThresholdAOne{} and is a prespecified stress rule rather than a
community-standard cutoff. Criterion \CriterionATwo{} requires that share to
be at most \ThresholdATwo{}, the necessary fit bound for an indivisible
component in the specified split geometry. Criterion \CriterionAThree{}
requires at least \ThresholdAThree{} effective components, incorporating the
specified reserve above the minimum number of groups. Criterion
\CriterionAFour{} requires largest-component share at most
$\ThresholdAFourFraction{}$, the equal-fold fit bound at the specified fold
count. As Section~\ref{sec:criteria} makes explicit, \StressCriterion{},
\CriterionATwo{}, and \CriterionAFour{} reuse the same statistic and are
neither independent tests nor a general theorem of split feasibility.

A source qualifies when both individual layers pass at least
\SingleLayerPassRequirement{} criteria and their union fails at least
\UnionFailRequirement{}. Gate~1 asks whether at least \QualifyingSources{}
gate-eligible sources meet this source-level rule. Because provenance can be
shared across derived datasets, outcomes are reported both by source and by
provenance family.

\subsection{Mechanism-discrimination and sensitivity boundaries}

Gate~2 asks whether a prespecified bridge-specific predictor explains the
qualifying pattern beyond simple graph density. The selected candidate,
\texttt{bridge\_edge\_density}, counts cross-container content edges per
container. It must separate qualifiers from nonqualifiers in the registered
direction, reach the registered association threshold with C5
largest-component share, and remain separated in the one-source-deletion
audit. Registered naive predictors are then tested as negative controls. In
particular, \texttt{mean\_union\_degree} measures ordinary C5 edge density.
The mechanism-discrimination requirement is not met if such a control also
satisfies the separation and outcome-association conditions. A repeated
source-level pattern and a bridge-specific explanation are therefore distinct
claims; establishing the former does not establish the latter.

The remaining analyses preserve narrower evidence domains. The registered
$\tau$ sweep concerns a discrete grid and a multiplicity-adjusted winner-flip
decision, not an equivalence test. The coverage-exposure and lexical audits
concern only the corpora and definitions represented in their source records.
These secondary checks bound interpretation of the main configuration
measurement; they do not enlarge the source-level rule or the population to
which its result applies.

\section{Union-induced collapse appears in \QualifyingSources{} of
\EligibleSources{} gate-eligible sources}
\label{sec:main-result}

Doc2Dial provides the clearest observed instance by collapse magnitude.  The
largest connected component contains \DocDialContentShare{} of units under
the content-only layer and \DocDialContainerShare{} under the container-only
layer.  Under transitive closure of their union, the corresponding share is
\DocDialUnionShare{}.  Only \DocDialUnionComponents{} components remain, and
the effective number of components falls to \DocDialUnionEffective{}.

MyFixit supplies a second, less extreme instance from a different provenance
family.  Its content-only and container-only largest-component shares are
\MyFixitContentShare{} and \MyFixitContainerShare{}, respectively; the union
share is \MyFixitUnionShare{}.  The union contains
\MyFixitUnionComponents{} components and has an effective component count of
\MyFixitUnionEffective{}.

Under the audit criteria defined in Section~\ref{sec:criteria}, MyFixit is the
only qualifying source whose individual layers both pass all
\RegisteredCriterionCount{} registered criteria.  Each Doc2Dial individual
layer fails only criterion
\StressCriterion{} and passes criteria \CriteriaATwoToAFour{}.

\begin{figure}[t]
  \centering
  \includegraphics[width=\linewidth]{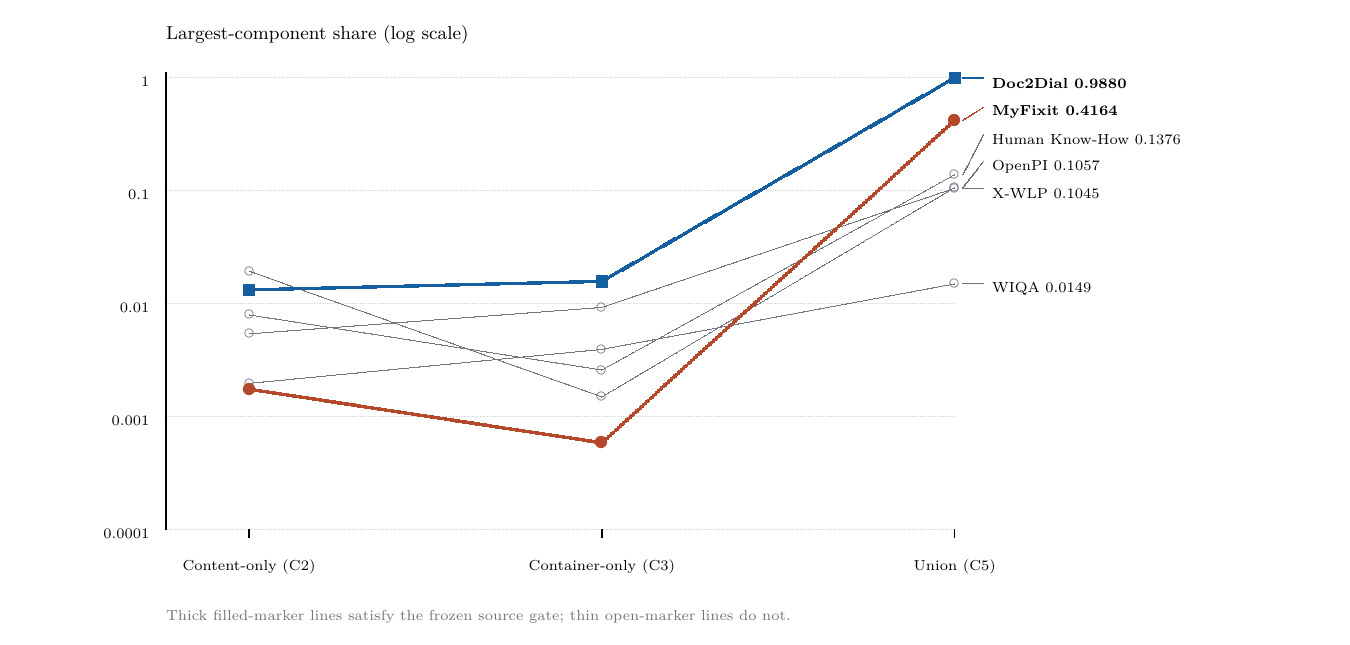}
  \caption{Largest-component share under the two individual layers and their
  union for all \EligibleSources{} gate-eligible sources.  The vertical axis is
  logarithmic.  Bold colored lines identify the \QualifyingSources{} sources satisfying the frozen source
  gate; gray lines identify nonqualifying sources.  The highlighting records
  an audit outcome, not a causal explanation.}
  \label{fig:section3-union-collapse}
\end{figure}

\QualifyingSources{} of the \EligibleSources{} gate-eligible sources in the
frozen panel satisfy the source Gate.  This \QualifyingRatio{} value is a
descriptive panel fraction.  It is not an estimate from independent trials or
a prevalence estimate.  MyFixit was the previously known qualifying source.
Among \CandidateSourcesScreened{} additionally screened candidates, Doc2Dial
was the sole qualifier (\AdditionalQualifyingSources{} of
\CandidateSourcesScreened{}).  The \QualifyingSources{} qualifying sources belong to
\QualifyingFamilies{} distinct provenance families,
\texttt{DOC2DIAL\_GOV} and \texttt{IFIXIT}.

WIQA provides a useful negative case: its union share is \WIQAUnionShare{},
only modestly above the \WIQAContentShare{} and \WIQAContainerShare{} shares
of its individual layers, and its content layer contains only
\WIQAContentEdges{} edges.  Union formation alone therefore does not entail
collapse; cross-container content repetition must be present in the measured
graph.

\begin{table}[t]
\centering
\caption{Closure metrics for the gate-eligible sources. C2 and C5 use the frozen near-duplicate threshold $\tau=0.85$; C3 uses exact container identity. $N_{\mathrm{eff}}=(\sum_j p_j^2)^{-1}$ for component unit shares $p_j$.}
\label{tab:section3-lattice}
\scriptsize
\setlength{\tabcolsep}{5pt}
\begin{tabular}{llrrr}
\toprule
Source & Layer & Components & Largest share & $N_{\mathrm{eff}}$ \\
\midrule
Doc2Dial & Content-only (C2) & 21,476 & 0.013132 & 2287.732 \\
 & Container-only (C3) & 488 & 0.015632 & 325.232 \\
 & Union (C5) & 18 & 0.988039 & 1.024 \\
\addlinespace[2pt]
MyFixit & Content-only (C2) & 84,777 & 0.001736 & 12870.890 \\
 & Container-only (C3) & 18,995 & 0.000582 & 10209.086 \\
 & Union (C5) & 5,375 & 0.416427 & 5.760 \\
\addlinespace[2pt]
Human Know-How & Content-only (C2) & 77,453 & 0.007897 & 12922.284 \\
 & Container-only (C3) & 9,967 & 0.002566 & 6014.607 \\
 & Union (C5) & 8,550 & 0.137558 & 52.215 \\
\addlinespace[2pt]
OpenPI & Content-only (C2) & 3,908 & 0.019259 & 1533.517 \\
 & Container-only (C3) & 810 & 0.001481 & 788.961 \\
 & Union (C5) & 683 & 0.105679 & 74.520 \\
\addlinespace[2pt]
WIQA & Content-only (C2) & 2,508 & 0.001955 & 2436.917 \\
 & Container-only (C3) & 379 & 0.003911 & 359.541 \\
 & Union (C5) & 341 & 0.014861 & 267.337 \\
\addlinespace[2pt]
X-WLP & Content-only (C2) & 2,213 & 0.005364 & 1422.480 \\
 & Container-only (C3) & 279 & 0.009195 & 226.169 \\
 & Union (C5) & 180 & 0.104470 & 49.318 \\
\bottomrule
\end{tabular}
\end{table}

Table~\ref{tab:section3-lattice} reports the content-only, container-only, and
union configurations for
every gate-eligible source.  The largest-component share captures the most
visible concentration, whereas the effective component count summarizes the
full component-size distribution.  In Doc2Dial, the union is close to a
single-component partition by both summaries; in MyFixit the concentration is
substantial but not complete.  The \NonqualifyingSources{} nonqualifying
sources remain in the table to make the \QualifyingRatio{} denominator and the negative cases
explicit.

\section{Audit criteria and observed rule-shape stability}
\label{sec:criteria}

The source-level rule comprises criteria \AllAuditCriteria{}. The
content-only and container-only layers must each pass at least
\SingleLayerPassRequirement{} criteria, while the union layer must fail at
least \UnionFailRequirement{}. This is the rule under which the
\QualifyingRatio{} result was obtained. Its thresholds, source eligibility,
and outcome are unchanged in the analysis below.

The criteria are operational components rather than independent statistical
tests. \StressCriterion{} is a prespecified stress criterion requiring
largest-component share at most \ThresholdAOne{}. \CriterionATwo{} applies
the same statistic at the specified smallest-split threshold
\ThresholdATwo{}; its grounding is the necessary fit of an indivisible
component. Lo-Hi illustrates that rationale by assigning connected components
independently while reporting that a giant component prevents its required
split; it does not supply the present A2 threshold
\cite{steshin2023_p4r24}. \CriterionAThree{} instead requires at least
\ThresholdAThree{} effective components. Its grounding combines the
GroupKFold requirement for at least the specified \RegisteredFoldCount{}
groups with a \AThreeReserveMultiplier{}-fold reserve; that reserve is a
stress margin, not an external theorem. \CriterionAFour{} returns to
largest-component share and requires at most $\ThresholdAFourFraction{}$,
the equal-fold fit threshold at the specified fold count. Once the fold count
$k$ is fixed, \CriterionAFour{} introduces no additional threshold-tuning
degree of freedom. Thus \StressCriterion{}, \CriterionATwo{}, and
\CriterionAFour{} are nested thresholds on one statistic, whereas
\CriterionAThree{} uses effective components. Counting criteria therefore
repeats weight on largest-component share.

In the observed \EligibleSources{}-source panel, every union layer fails
\StressCriterion{}; all \AllUnionAOneFailures{} union outcomes have that
status. Consequently, within this panel only, the union requirement is
equivalent to failing at least \AdditionalUnionFailuresNeeded{} further
criteria among \CriteriaATwoToAFour{}. This is a panel-specific equivalence in
the observed outcomes, not a permanent rewrite of the decision rule.

\begin{center}
\centering
\captionsetup{hypcap=false}
\captionof{table}{Independently recomputed union-layer criterion outcomes within the frozen 6-source panel. The failure count uses A2--A4 only; qualifying status remains the original A1--A4 source Gate.}
\label{tab:section4-union-criteria}
\small
\setlength{\tabcolsep}{4pt}
\begin{tabular}{l@{\hspace{10pt}}cccccc}
\toprule
Source & A1 & A2 & A3 & A4 & A2--A4 failures & Frozen qualifier \\
\midrule
Human Know-How & FAIL & PASS & PASS & PASS & 0/3 & no \\
MyFixit & FAIL & FAIL & FAIL & FAIL & 3/3 & yes \\
OpenPI & FAIL & PASS & PASS & PASS & 0/3 & no \\
WIQA & FAIL & PASS & PASS & PASS & 0/3 & no \\
Doc2Dial & FAIL & FAIL & FAIL & FAIL & 3/3 & yes \\
X-WLP & FAIL & PASS & PASS & PASS & 0/3 & no \\
\bottomrule
\end{tabular}
\end{center}

Table~\ref{tab:section4-union-criteria} shows the independently recomputed
union-layer matrix. Within this \EligibleSources{}-source panel, the
\QualifyingSources{} qualifiers have the \QualifierATwoToAFourPattern{}
failure pattern on \CriteriaATwoToAFour{}, whereas all
\NonqualifyingSources{} nonqualifiers have the
\NonqualifierATwoToAFourPattern{} pattern. No source has an intermediate
failure count; the observed count of such sources is
\IntermediatePatternCount{}. This complete separation is consistent with the
\QualifyingRatio{} qualifying set; it is not a significance test, predictive
validation, or an additional source of replication evidence.

As a rule-shape check, all \NonzeroCutoffCount{} nonzero failure-count cutoffs
\CutoffOneOfThree{}, \CutoffTwoOfThree{}, and \CutoffThreeOfThree{} each
select the same sources as the source-level rule. The
\SingleCriterionCheckCount{} single-criterion checks using
\CriterionATwo{}, \CriterionAThree{}, or \CriterionAFour{} also select that
same set within this panel. The \CriterionAFour{} result is the cleanest
single-criterion check here because, after the fold count is fixed, it adds no
further threshold choice. These checks do not show that criterion choice is
insensitive outside this panel.

This \CriteriaATwoToAFour{} analysis is post hoc, presentation-only, and
descriptive. It neither changes the original \AllAuditCriteria{} rule nor
defines a new prespecified decision rule. The separation does not establish a
mechanism and is not evidence about prevalence or behavior outside the fixed
panel.

\section{The bridging mechanism is not identifiable in the six-source panel}
\label{sec:mechanism-identifiability}

The prespecified bridging predictor, \texttt{bridge\_edge\_density}, shows a
strong association in the fixed panel: it perfectly separates the
\MechanismQualifiers{} qualifying sources from the remaining sources, meets
the registered correlation condition, and has a strictly separated
full-panel range (minimum qualifier \BridgeQualifierMinimum{} versus maximum
nonqualifier \BridgeNonqualifierMaximum{}). Nevertheless, the
mechanism-discrimination criterion is not met. The registered naive control,
\texttt{mean\_union\_degree}, completely separates the two qualifiers from
the four nonqualifiers. Its Spearman correlation with C5 largest-component
share is $\rho_s=\NaiveOutcomeRho{}$, meeting the registered
$\rho_s\geq\RegisteredOutcomeRhoThreshold{}$ condition. The negative control
therefore satisfies both the separation and outcome-correlation conditions.

Under the frozen Gate-2 rule, this result is attributable to lack of
discriminant validity against the registered negative control, not to absence
of an observed bridge association in this six-source panel. Across the
\MechanismPanelSources{} sources, the predictor vectors have Spearman rank
correlation $\rho_s=\PredictorCrossRho{}$. Their descending ranks agree for
every source except OpenPI and WIQA, which exchange adjacent positions; there
is only \PredictorDiscordantPairs{} discordant source pair. Both predictors
place Doc2Dial and MyFixit first and second, so the observed binary outcome
cannot distinguish bridge-specific structure from simple union-graph density.

\begin{center}
\begin{minipage}{0.86\linewidth}
\centering
\captionsetup{hypcap=false}
\captionof{table}{Frozen predictor values and within-panel descending ranks. The two predictors differ only in the OpenPI/WIQA ordering.}
\label{tab:section5-source-ranks}
\small
\setlength{\tabcolsep}{4pt}
\begin{tabular}{lcrr}
\toprule
Source & Qualifies & Bridge density (rank) & Mean union degree (rank) \\
\midrule
Human Know-How & no & 20.891 (3) & 17.536 (3) \\
MyFixit & yes & 106.648 (2) & 39.273 (2) \\
OpenPI & no & 4.100 (5) & 5.773 (6) \\
WIQA & no & 0.161 (6) & 6.160 (5) \\
Doc2Dial & yes & 405.340 (1) & 108.686 (1) \\
X-WLP & no & 6.864 (4) & 17.288 (4) \\
\bottomrule
\end{tabular}
\end{minipage}
\end{center}

Table~\ref{tab:section5-predictor-summary} records both predictors' frozen
separation thresholds, slacks, and correlations with C5 largest-component
share. Three correlations must be kept distinct: bridge-to-C5 is
$\BridgeOutcomeRho{}$; negative-control-to-C5 is $\NaiveOutcomeRho{}$; and
the cross-predictor correlation is $\PredictorCrossRho{}$. The
negative-control-to-C5 value meets the registered
$\rho_s\geq\RegisteredOutcomeRhoThreshold{}$ condition. None of the
differences between these outcome correlations, thresholds, or slacks is
credited in favor of the bridge predictor. Reporting the values is an audit
record, not a post hoc comparison that rescues the mechanism.

\begin{center}
\centering
\captionsetup{hypcap=false}
\captionof{table}{Frozen separation records for the selected bridge predictor
and the registered naive control. Threshold, slack, and outcome correlation
are reported for audit completeness; under the frozen negative-control rule,
their numerical differences do not support the bridge predictor.}
\label{tab:section5-predictor-summary}
\small
\setlength{\tabcolsep}{4pt}
\begin{tabular}{lcccc}
\toprule
Predictor & Separates & Threshold & Slack & $\rho_s$ with C5 share \\
\midrule
\texttt{bridge\_edge\_density} & yes & \BridgeSeparationThreshold{} &
\BridgeSeparationSlack{} & \BridgeOutcomeRho{} \\
\texttt{mean\_union\_degree} & yes & \NaiveSeparationThreshold{} &
\NaiveSeparationSlack{} & \NaiveOutcomeRho{} \\
\bottomrule
\end{tabular}
\end{center}

Table~\ref{tab:section5-deletion-audit} is a one-source-deletion separability
audit. For each row, the threshold is recomputed on the retained sources and
the omitted source is not classified; this is not predictive cross-validation.
Because full-panel separation is strict
(\BridgeQualifierMinimum{} $>$ \BridgeNonqualifierMaximum{}) and both classes
remain after every single deletion, persistence under all
\DeletionAuditRows{} deletions is mechanically implied. The table verifies
the arithmetic of that implication but supplies no independent robustness
evidence and cannot change the mechanism conclusion.

\begin{center}
\begin{minipage}{0.86\linewidth}
\centering
\captionsetup{hypcap=false}
\captionof{table}{One-source-deletion separability audit for the registered bridge predictor. For each row, the threshold is recomputed on the retained sources and the omitted source is not classified. This is not predictive cross-validation. Because the full panel is strictly separated and both classes remain after every deletion, persistence is mechanically implied and supplies no independent robustness evidence.}
\label{tab:section5-deletion-audit}
\small
\setlength{\tabcolsep}{5pt}
\begin{tabular}{lccc}
\toprule
Omitted source & Separates & Threshold & Slack \\
\midrule
Human Know-How & yes & 56.7560 & 49.8922 \\
MyFixit & yes & 213.1158 & 192.2244 \\
OpenPI & yes & 63.7698 & 42.8783 \\
WIQA & yes & 63.7698 & 42.8783 \\
Doc2Dial & yes & 63.7698 & 42.8783 \\
X-WLP & yes & 63.7698 & 42.8783 \\
\bottomrule
\end{tabular}
\end{minipage}
\end{center}

A key limitation of the mechanism design is that predictor distinguishability
was not assessed before the Gate-2 decision rule was frozen. Freezing the rule
before outcome interpretation prevented a post hoc rescue, but it did not
make the control orthogonal. The analysis shows that unique explanatory gain
was not demonstrated; it cannot decide which of two nearly aligned predictor
families better accounts for the observed association. A future result-blind
predictor-geometry audit should precede outcome testing, and the source panel
should contain cases where the candidate mechanism and simple graph density
make different predictions.

The observed correlation shows that the predictor rankings are nearly aligned
in this six-source panel, but no defensible numerical panel-size target can be
derived. Without a source-sampling design, a fitted outcome model, or a defined
precision criterion, the required number of sources remains unidentified.

The density control also depends on how container connectivity is represented.
C3 expands each container into a clique, so connected components depend on
container membership but \texttt{mean\_union\_degree} depends on the edge
expansion. A star or spanning forest can preserve the same within-container
connectivity with fewer edges. In a post hoc, panel-specific sensitivity
analysis using only frozen aggregate counts, we set $T_s=n_s-c_s$ and
$O_s=E_{2,s}+E_{3,s}-E_{5,s}$ and bound the unknown overlap $x_s$ between C2
and forest edges by
$\max(0,T_s-(E_{3,s}-O_s))\leq x_s\leq\min(T_s,O_s)$; the corresponding union
edge count is $E_{2,s}+T_s-x_s$. The conservative aggregate-compatible
mean-degree intervals leave the lowest qualifier value
(\EditorialForestQualifierLowerBound{}) above the highest nonqualifier value
(\EditorialForestNonqualifierUpperBound{}). Enumerating every compatible weak
rank order, with average ranks for permitted ties, gives correlations with C5
largest-component share from \EditorialForestSpearmanMinimum{} to
\EditorialForestSpearmanMaximum{}, all above the registered
\RegisteredOutcomeRhoThreshold{} threshold. These bounds do not identify a
unique spanning tree, and their endpoints need not both be attained by a
constructible forest; graph-connectivity constraints could narrow them. No raw
graph was read or concrete forest constructed. The original control remains
clique-based, mean degree is not representation-invariant, and the frozen
Gate-2 result is unchanged.

\section{The registered \texorpdfstring{$\tau$}{tau} grid is decision-inert, not metric-invariant}
\label{sec:tau-inertness}

The frozen sweep artifact covers \TauSweepSources{} sources, MyFixit and
X-WLP, at \TauGridPoints{} registered discrete settings from exact matching
through $\tau=\TauGridLowerBound{}$.  It is not a continuous-interval study
and it is not a six-source sweep.  Accordingly, this section reports only the
quantities present in the two permitted frozen records.

The raw component structure is not invariant.  Of the
\TauRawAuditedMetricCount{} audited quantities---component count,
largest-component share, and effective components---
\TauRawInvariantMetricCount{} remain constant across the registered grid for
both sources.  Table~\ref{tab:section6-tau-endpoints} shows the endpoint
values.  Component counts and effective-component counts decline, while
largest-component share moves at the lower grid points.  Calling these raw
statistics unchanged would therefore contradict the frozen sweep.

\begin{center}
\begin{minipage}{0.98\linewidth}
\centering
\captionsetup{hypcap=false}
\captionof{table}{Endpoint audit of the registered discrete $\tau$ grid. Shares are generated as exact component size divided by frozen source units, not reused as ratio inputs after display formatting. Raw component statistics change between exact matching and $\tau=0.50$; the final column reports the sum of Holm-surviving flips over all frozen fixture rows and grid points for each source.}
\label{tab:section6-tau-endpoints}
\small
\setlength{\tabcolsep}{3pt}
\begin{tabular}{lrrrrrrr}
\toprule
Source & \multicolumn{2}{c}{$n_{\rm comp}$} & \multicolumn{2}{c}{top1 share} & \multicolumn{2}{c}{$N_{\rm eff}$} & Holm flips \\
 & exact & 0.50 & exact & 0.50 & exact & 0.50 & all rows \\
\midrule
MyFixit & 85,423 & 78,563 & 0.001736 & 0.007001 & 13292.7542 & 6279.5639 & 0 \\
X-WLP & 2,222 & 2,083 & 0.005364 & 0.005875 & 1448.5611 & 1204.8758 & 0 \\
\bottomrule
\end{tabular}
\end{minipage}
\end{center}

The decision-level negative result is narrower.  The frozen winner-identity
record contains \TauWinnerRows{} fixture--source--grid rows.  It records
\TauRawFlips{} raw flips but \TauHolmSurvivingFlips{} Holm-surviving flips,
and its surviving-flips table is empty.  Thus ``$\tau$ inertness'' here means
that the registered multiplicity-adjusted flip decision does not change on
the registered discrete grid.  It does not mean that retained-query counts,
rank correlations, winner tie sets, or component statistics are identical.
The absence of Holm-surviving flips is a negative result under the registered
decision rule; it is not an equivalence test and does not establish that
$\tau$ has zero effect.

Neither permitted source records a six-source qualifier-set trajectory over
this grid.  We therefore do not claim that the frozen qualifier set is
invariant from exact matching through $\tau=\TauGridLowerBound{}$.  Such a
claim would require a corresponding result-blind six-source sweep artifact,
which is absent from the frozen record used here.

The edge-family contrast is separately visible in the frozen lattice.
Table~\ref{tab:section6-edge-family} compares the near-duplicate text layer
alone (C2) with its union with source-group edges (C5).  For both recorded
sources, the configurations have different component counts,
largest-component shares, and effective-component counts.  This is an
observed configuration contrast; it does not identify a general causal
mechanism.

\begin{center}
\begin{minipage}{0.94\linewidth}
\centering
\captionsetup{hypcap=false}
\captionof{table}{Frozen near-duplicate edge-family comparison. C2 uses text-near edges alone; C5 uses their union with source-group edges. These are observed configuration contrasts, not a general causal claim.}
\label{tab:section6-edge-family}
\small
\setlength{\tabcolsep}{4pt}
\begin{tabular}{lrrrrrr}
\toprule
Source & \multicolumn{2}{c}{$n_{\rm comp}$} & \multicolumn{2}{c}{top1 share} & \multicolumn{2}{c}{$N_{\rm eff}$} \\
 & C2 & C5 & C2 & C5 & C2 & C5 \\
\midrule
MyFixit & 84,777 & 5,375 & 0.001736 & 0.416427 & 12870.890 & 5.760 \\
X-WLP & 2,213 & 180 & 0.005364 & 0.104470 & 1422.480 & 49.318 \\
\bottomrule
\end{tabular}
\end{minipage}
\end{center}

On the one outcome reported in both tables, MyFixit's largest-component share,
the exact-to-$\tau=\TauGridLowerBound{}$ endpoint ratio is
\MyFixitTauTopOneRatio{}, whereas the C2-to-C5 ratio is
\MyFixitEdgeTopOneRatio{}; the latter divided by the former is
\MyFixitTopOneRatioQuotient{}.  This is a descriptive magnitude comparison on
a single statistic, not an effect-size test or a significance comparison, and
it is not extrapolated beyond the registered grid or this frozen panel.  All
three ratios are computed from exact integer component sizes rather than the
displayed shares.  The Gate-C exact endpoint and the lattice C2 configuration
happen to share that size.  The same largest-component size does not make the
two configurations identical.

Within this frozen design, C2 and C5 differ substantially in the audited
component summaries, whereas the registered $\tau$ sweep yields no
Holm-surviving winner flips.  These observations concern different outcome
levels and are not interpreted as a direct effect-size comparison.  The C2/C5
difference is an observed configuration contrast without a general causal
interpretation.  It does not imply that edge family is universally decisive.
It does not imply that a near-duplicate threshold is generally unimportant,
that values outside the registered grid are inert, or that the result
transfers to other corpora or other near-duplicate methods.

\section{A coverage-exposure near-identity}
\label{sec:coverage-exposure}

The registered naive-negative exposure is the fraction of all pairs that a
naive policy would call negative but that the source policy leaves unresolved.
For each corpus, let $U$ be the number of unresolved pairs, $R$ the number of
licensed refuted pairs, $I$ the number of incomplete units, and $N$ the total
number of units.  The two quantities compared here are therefore $U/(R+U)$ and
$I/N$.

For MyFixit, the source-recorded counts give
\[
  \frac{\MyFixitCoverageUnresolvedPairs{}}
       {\MyFixitCoverageNaivePairs{}}
  = \MyFixitCoverageExposure{},
  \qquad
  \frac{\MyFixitCoverageIncompleteUnits{}}
       {\MyFixitCoverageUnits{}}
  = \MyFixitCoverageIncompleteFraction{},
\]
an absolute difference of \MyFixitCoverageFractionGap{}, or
\MyFixitCoveragePercentagePointGap{} percentage points.  For X-WLP, the same
arithmetic gives
\[
  \frac{\XWLPCoverageUnresolvedPairs{}}
       {\XWLPCoverageNaivePairs{}}
  = \XWLPCoverageExposure{},
  \qquad
  \frac{\XWLPCoverageIncompleteUnits{}}
       {\XWLPCoverageUnits{}}
  = \XWLPCoverageIncompleteFraction{},
\]
an absolute difference of \XWLPCoverageFractionGap{}, or
\XWLPCoveragePercentagePointGap{} percentage points.

\begin{center}
\begin{minipage}{\linewidth}
\centering
\captionsetup{hypcap=false}
\captionof{table}{Coverage exposure and the unit-level incompleteness fraction under the one registered exposure definition.  Gap is the absolute difference in percentage points; the final columns show the corresponding candidate-pair mass per complete and incomplete unit.}
\label{tab:section7-coverage-exposure}
\small
\setlength{\tabcolsep}{4pt}
\begin{tabular}{lrrrrr}
\toprule
Corpus & Exposure & Incomplete & Gap (pp) & Pairs/complete & Pairs/incomplete \\
\midrule
MyFixit & \MyFixitCoverageExposure{} & \MyFixitCoverageIncompleteFraction{} & \MyFixitCoveragePercentagePointGap{} & \MyFixitCoverageCompletePairsPerUnit{} & \MyFixitCoverageIncompletePairsPerUnit{} \\
X-WLP & \XWLPCoverageExposure{} & \XWLPCoverageIncompleteFraction{} & \XWLPCoveragePercentagePointGap{} & \XWLPCoverageCompletePairsPerUnit{} & \XWLPCoverageIncompletePairsPerUnit{} \\
\bottomrule
\end{tabular}
\end{minipage}
\end{center}

The near identity follows from the pair weighting in these two records.  The
candidate-universe mass is \MyFixitCoverageCompletePairsPerUnit{} licensed
refuted pairs per complete MyFixit unit and
\MyFixitCoverageIncompletePairsPerUnit{} unresolved pairs per incomplete
unit.  The corresponding X-WLP values are
\XWLPCoverageCompletePairsPerUnit{} and
\XWLPCoverageIncompletePairsPerUnit{}.  Because those complete- and
incomplete-unit weights are close within each corpus, weighting units by their
candidate-pair mass changes the incompleteness fraction only slightly.

This is an arithmetic near-identity observed on \CoverageExposureCorpora{}
corpora under one exposure definition.  It does not establish the same
relationship for another exposure definition or another corpus, and it does
not establish that coverage-exposure statistics are uninformative beyond
these two cases.  On these two corpora, the registered statistic carries
approximately the same information as the annotated-field completeness rate.
This observation is not a critique of any published work.

\section{Structured-field unresolved status does not certify lexical silence}
\label{sec:structured-lexical-silence}

E1 assigns \texttt{UNRESOLVED} when a structured field lies outside the
frozen completeness policy or cannot license negative entailment.  That
open-world status prevents an unsupported negative label; it is not a
certificate that attached natural-language text lacks relevant vocabulary.
The distinction is evaluated here only as a deterministic lexical lower-bound
diagnostic over the \LexicalAuditCorpora{} corpora represented in the sole
permitted aggregate report.

\begin{center}
\begin{minipage}{\linewidth}
\centering
\captionsetup{hypcap=false}
\captionof{table}{Deterministic lexical lower-bound diagnostic reproduced arithmetically from the hash-verified aggregate report. Positive counts identify frozen-vocabulary cues; they are not semantic labels or requested-value-specific mention rates.}
\label{tab:section8-lexical-diagnostics}
\small
\setlength{\tabcolsep}{3pt}
\begin{tabular}{p{0.12\linewidth}p{0.46\linewidth}rrr}
\toprule
Corpus & Diagnostic & Unresolved & Positive & Rate \\
\midrule
MyFixit & Step text: exact normalized frozen-tool cue & \MyFixitUnresolvedUnits{} & \MyFixitStepCueCount{} & \MyFixitStepCueRate{} \\
MyFixit & Guide title, subject, or toolbox: exact normalized frozen-tool cue & \MyFixitUnresolvedUnits{} & \MyFixitContextCueCount{} & \MyFixitContextCueRate{} \\
MyFixit & Structured tools field intersects frozen vocabulary & \MyFixitUnresolvedUnits{} & \MyFixitToolsFieldCueCount{} & \MyFixitToolsFieldCueRate{} \\
X-WLP & Operation sentence: frozen location-family lexical cue & \XWLPUnresolvedOperations{} & \XWLPSentenceCueCount{} & \XWLPSentenceCueRate{} \\
\bottomrule
\end{tabular}
\end{minipage}
\end{center}

The MyFixit step-text predicate is ``contains at least one exact normalized
cue from the frozen tool vocabulary.''  Under that predicate,
\MyFixitStepCueCount{} of
\MyFixitUnresolvedUnits{} unresolved units (\MyFixitStepCueRate{}) contain at
least one exact normalized cue from the frozen tool vocabulary.  The broader
guide-title, subject, or toolbox context produces \MyFixitContextCueCount{}
(\MyFixitContextCueRate{}), while the structured tools field produces
\MyFixitToolsFieldCueCount{} (\MyFixitToolsFieldCueRate{}).  For X-WLP,
\XWLPSentenceCueCount{} of \XWLPUnresolvedOperations{} unresolved operations
(\XWLPSentenceCueRate{}) have an operation sentence that contains a frozen
location-family lexical cue.  Reporting every frozen diagnostic prevents the
broad-context rate from standing alone.

Any positive count is enough to reject the blanket implication that E1
\texttt{UNRESOLVED} entails absence of every frozen-vocabulary lexical cue.
The converse does not follow.  These matches are not semantic labels; a cue
need not establish the exact requested relation, and a missing cue does not
certify natural-language or semantic silence.  Thus structured-field
unresolved status does not certify lexical silence, but this audit does not
establish semantic sufficiency.

The aggregate report also records requested-value degeneracy.  Every MyFixit
unresolved replay assertion uses the same \RequestedUniverseIndex{} tool, and
every X-WLP unresolved replay assertion uses the same
\RequestedUniverseIndex{} location family.  The displayed rates therefore
cannot be interpreted as requested-value-specific mention rates or as
coverage over a representative requested-value distribution.

This section is an aggregate-report arithmetic reproduction from one
hash-verified source document, not a raw-corpus independent recount or an
end-to-end reproduction.  Its scope is limited to these corpora, the frozen E1
construction, and the two recorded relation vocabularies.  It does not extend
to other corpora, relations, fields, or unresolved definitions, and it is not
a critique of published work.

\section{Related work and position of this measurement}
\label{sec:related-work}

\subsection{Split feasibility and leakage-aware partitioning}

Similarity-aware and leakage-reduced partitioning form a mature line of work
in biomolecular machine learning.  Lo-Hi shows how a giant component in a
single molecular-similarity graph can prevent its required split and formulates
constrained vertex removal to recover a feasible partition
\cite{steshin2023_p4r24}.  GraphPart instead uses restricted single-linkage
partitioning over a sequence-homology graph, with iterative sequence movement
or removal to trade retention against separation
\cite{teufel2023_graphpart}.  Refnd constructs a thresholded proximity graph,
splits at connected components, and optionally subdivides dense components
with Leiden before post-filtering \cite{lavertu2026_refnd}.  These works give
distinct graph-based responses to split feasibility; they should not be read
as instances of one identical construction.

Other methods combine several similarity constraints without performing the
present layer-versus-union audit.  DataSAIL clusters typed entities and
optimizes their fold assignments to reduce cross-fold similarity; conflicting
assignments in a two-dimensional setting can require interaction removal
\cite{joeres2025_p4r25}.  LP-PDBBind jointly controls protein-sequence and
ligand similarity across train, validation, and test partitions and discards
remaining cross-category conflicts before retraining scoring functions
\cite{li2026_lppdbbind}.  PLINDER constructs metric-specific protein, pocket,
protein--ligand-interaction, and ligand graphs, records their connected
components and Louvain communities, and permits split configurations that use
several graphs and neighbor depths \cite{durairaj2024_plinder}.  Together,
these studies establish a substantial prior literature on similarity-aware
partition design and its retention costs.

\subsection{Transitive closure, entity matching, and component formation}

The propagation risk of closure also predates the present measurement.  In
record linkage, Gu et al. describe combining independent matching passes by
transitive closure and warn that a false-positive match can propagate across
passes \cite{gu2003_p4r06}.  In knowledge-graph entity matching, Baas,
Dastani, and Feelders compare closure with weighted cluster editing, which adds
and deletes candidate links to enforce transitivity
\cite{baas2021_p4r11}.  This literature supplies the relevant closure
and repair background, but its evaluated outcome is entity-matching quality
rather than component-disjoint benchmark construction across a corpus panel.
Classical random-graph theory derives giant-component phase-transition
conditions for arbitrary-degree models, including bipartite graphs
\cite{newman2001_p4r01}.  We cite this result only as background on
giant-component formation; the deterministic typed-layer union audited here is
not a test of that random-graph model.

\subsection{Near-duplicate and incomplete-judgment evaluation}

Information-retrieval evaluation provides a separate adjacent line.  Froebe et
al. reproduce and extend a study of content-equivalent documents across TREC
Terabyte, Web, and Core tracks, showing that duplicate handling can change
system evaluation \cite{froebe2020_nearduplicates}.  Buckley and Voorhees
introduce bpref to compare retrieval systems when relevance judgments are
incomplete \cite{buckley2004_bpref}.  We use these works to delimit evaluation
concerns around redundancy and missing judgments; neither is evidence for a
typed relation-union split construction.

\subsection{Position of the cross-domain measurement}

The central direct comparator identified in the bounded search is Guvenilir
and Doğan's DTI study \cite{guvenilir2023_dti}.  In that setting, the authors
construct a heterogeneous network from
\SectionNineComparatorRelationTypes{} relation types---compound similarity,
protein similarity, and compound--target interactions---compute connected
components, and report that a giant component prevents component-disjoint
splitting.  Their response uses Louvain-guided edge and node deletion.  The
reported study covers \SectionNineComparatorDatasets{} protein-family
datasets, \SectionNineComparatorSplitVariants{} split variants per family,
and \SectionNineComparatorModels{} trained models.  We therefore attribute
the merged-relation, giant-component, split-infeasibility, and
deletion/community-repair chain to that work.

That ownership boundary is reinforced by other partial direct precedents.  The PPI
leakage study groups a graph whose edges encode observed protein interactions
or shared sequence clusters before measuring residual interface-structure
leakage \cite{bushuiev2024_ppi}; PLINDER supports multi-graph split
configurations \cite{durairaj2024_plinder}.  Among the works reviewed here, we
did not identify in the reported designs the conjunction used by this study:
one common operational Gate applied symmetrically to individual edge families
and their union, individual-layer passes followed by union failure, and a
frozen cross-domain procedural-corpus panel.  This is a bounded-search
positioning statement, not an assertion about unreviewed literature.

The narrower object of the present measurement is therefore the symmetric
layer/union audit and its frozen panel outcome.  The
\SectionNineDescriptivePanelFraction{} result is a descriptive panel fraction,
not a population-frequency estimate.  The analysis also reports
component-based split-capacity indicators and retains the registered negative
result from Section~\ref{sec:mechanism-identifiability}: the bridge-specific
predictor was not distinguished from the union-density control.  This paper
contributes neither a splitting algorithm nor an entity-resolution or
Louvain/community repair method; it offers no giant-component theorem or
causal validation of bridging.  We also do not claim priority for observing
that merged relations can produce a giant component that obstructs a
component-disjoint split.

\section{Limitations and scope}
\label{sec:limitations}

\subsection{Panel and operational-rule scope}

The evaluation panel is small, outcome-enriched, and not drawn at random from
a sampling frame for procedural corpora. MyFixit was already known to satisfy
the source-level rule, while the remaining \CandidateSourcesScreened{}
sources were subsequently screened candidates. Doc2Dial was the only
additional qualifier found in that screening. The resulting
\QualifyingRatio{} value is therefore a descriptive fraction of this
\EligibleSources{}-source panel, not an estimate from independent trials or a
prevalence estimate. The panel spans several provenance families and domains,
but neither its composition nor its selection process supports a
population-level frequency claim.

Eligibility is structural rather than a claim of genre homogeneity. Doc2Dial
contains task-oriented government-service documents, whereas other panel
members include repair guides, instructional articles, procedure paragraphs,
and laboratory protocols. It meets the formal unit, container, and content-
field requirements, but that does not make the sources genre-equivalent.
Panel heterogeneity therefore limits external interpretation without changing
the fixed eligibility decision.

The requirement of at least two qualifying sources is an operational
replication rule rather than a statistical replication criterion. Because
MyFixit was the known qualifier, its practical application required at least
one further qualifier among the five screened candidates. Doc2Dial met that
additional-source requirement. This is neither a set of independent repeated
trials nor a basis for population-frequency inference.

The source-level decision rule is likewise an operational audit, not a general
theorem of split feasibility. \StressCriterion{}, \CriterionATwo{}, and
\CriterionAFour{} are nested thresholds on largest-component share, whereas
\CriterionAThree{} uses effective components. Their count therefore weights
largest-component concentration repeatedly; the
\RegisteredCriterionCount{} outcomes are not statistically independent. The
reserve behind \CriterionAThree{} is a prespecified stress margin, not an
externally established constant. Moreover, the complete separation produced
by \CriteriaATwoToAFour{} in this panel is a post hoc presentation check: it
leaves the original \AllAuditCriteria{} rule unchanged and does not show that
the same indicators are necessary or sufficient for useful splits elsewhere.

Simple container-size summaries do not reproduce the qualifying set. X-WLP
has a larger mean container size than MyFixit
(\EditorialXWLPMeanUnitsPerContainer{} versus
\EditorialMyFixitMeanUnitsPerContainer{}) but does not qualify. Human
Know-How has a larger maximum container size than MyFixit
(\EditorialHumanKnowHowMaximumUnitsPerContainer{} versus
\EditorialMyFixitMaximumUnitsPerContainer{}) and also does not qualify.
Doc2Dial's mean and maximum container sizes
(\EditorialDocTwoDialMeanUnitsPerContainer{} and
\EditorialDocTwoDialMaximumUnitsPerContainer{}) are comparatively large, so
container scale may contribute to its collapse magnitude. The narrower result
is only that these simple summaries, by themselves, do not separate both
qualifiers from all nonqualifiers; it does not establish that container size
is irrelevant or cannot contribute.

\subsection{Mechanism and sensitivity limits}

The observed association does not identify a bridge-specific mechanism.
\texttt{bridge\_edge\_density} and the registered union-density control share
union-edge information and are strongly rank-aligned in the
\MechanismPanelSources{}-source panel. Because the control satisfies the
registered separation and outcome-correlation conditions, the
mechanism-discrimination requirement is not met. The one-source-deletion
separability pattern adds no predictive evidence: strict full-panel
separation and retention of both outcome classes make that persistence
mechanical. A stronger design would audit predictor geometry before observing
outcomes and recruit result-blind sources on which the candidate mechanism and
a density control make discordant predictions. The present panel supplies
neither a defensible required sample size nor evidence that simple density is
itself causal.

The density control is not representation-invariant, although a post hoc
spanning-forest sensitivity bound yields the same qualitative negative-control
outcome within this panel. The original control is defined on the C3 clique
expansion and remains unchanged. The aggregate-compatible bounds in
Section~\ref{sec:mechanism-identifiability} preserve qualifier/nonqualifier
separation and keep every compatible rank correlation above the registered
threshold, but they are panel-specific conservative intervals rather than
exact values for a chosen forest. They neither prove representation invariance
nor reopen the frozen mechanism decision.

The secondary analyses constrain other interpretations but have narrower
domains. Section~\ref{sec:tau-inertness} covers \TauSweepSources{} sources and
\TauGridPoints{} registered discrete settings from exact matching through
$\tau=\TauGridLowerBound{}$; it is neither a continuous-interval analysis nor
a six-source qualifier sweep. Raw component summaries change, while
\TauHolmSurvivingFlips{} winner flips survive the registered Holm decision, so
decision-level inertness is not metric invariance or an equivalence test.
Section~\ref{sec:coverage-exposure} covers \CoverageExposureCorpora{} corpora
under one exposure definition, where pair weighting makes exposure
approximately re-express annotated-field incompleteness; it does not extend
that arithmetic relation to other definitions or corpora.
Section~\ref{sec:structured-lexical-silence} also covers only
\LexicalAuditCorpora{} corpora and reproduces arithmetic from one frozen
aggregate report rather than recounting the raw corpora. Its lexical cues are
not semantic labels, cue absence does not establish semantic silence, and
requested-value degeneracy limits the reported rates. Together these audits
bound the findings rather than establishing general threshold, exposure, or
vocabulary behavior.

\subsection{Evidence and reproducibility boundary}

The numerical claims are supported by frozen aggregate artifacts, and
separate consistency checks connect those records to the generated tables and
manuscript. That closure is not a complete raw-data-to-paper reproduction. The
recovered project lacks the complete raw corpora, inherited protocol tree,
locked environment, and a one-command path from raw inputs to the reported
panel; consequently, the current package supports inspection and
aggregate-level recomputation rather than a trusted public runner for the
original pipeline.

One unresolved parser risk illustrates the distinction. The recovered Human
Know-How loader passes UTF-8 Turtle label bytes through
\texttt{unicode\_escape}, which can introduce mojibake before text
normalization and thereby affect content-near-duplicate edges and C2/C5
component structure. No corrected-parser rerun is available, so the direction
and magnitude of any change to the frozen Human Know-How metrics cannot be
quantified. Current evidence does not require changing the frozen qualifier
set, but it also cannot establish that a corrected parser would leave those
metrics or that set unchanged. Raw-data reproduction and release of a trusted
runner therefore remain separate from the aggregate evidence used here.

\subsection{Prior-art and contribution scope}

The bounded search is not exhaustive literature proof, implies no criticism
of any research community, and cannot establish the absence of an unreviewed
comparator. Guvenilir and Doğan previously established the merged-relation,
giant-component, split-infeasibility, and deletion/community-repair chain
reviewed in Section~\ref{sec:related-work}. This paper offers neither a
first-observation claim nor a new split algorithm, Louvain/community repair
method, giant-component theorem, or causal bridge mechanism. Its retained
object is the symmetric comparison of typed layers with their union across a
fixed cross-domain procedural-corpus panel, together with negative findings
and explicit evidence limits. That measurement remains interpretable within
the stated panel and operational definitions but does not establish broader
generality.

\section{Reproducibility and artifact scope}
\label{sec:reproducibility}

\subsection{Separated verification objects}

The public materials separate two verification objects. The companion
aggregate artifact contains the frozen records supporting the reported
measurements, the scripts used to regenerate the scientific tables and
Figure~\ref{fig:section3-union-collapse}, and separate consistency checks for
those aggregate-derived outputs. Existing integrity records apply to those
frozen records and generated scientific assets. They do not automatically
cover the later Abstract, Introduction, Conclusion, or other editorial prose.

The final manuscript source package contains the final TeX source,
bibliography, compilation figures and tables, and the panel-summary and
representation-sensitivity CSV files with their deterministic generation
script. It is checked through clean extraction, dependency presence, citation
resolution, compilation, derived-file byte-for-byte comparison, and PDF font
preflight rather than claimed to be covered by the earlier aggregate
integrity records. The manuscript-level compile check is distinct from the
existing byte-identity records for the frozen aggregate evidence.

When the manuscript source package is paired with the companion aggregate
artifact, a clean-directory check rebuilds the aggregate-derived outputs, the
editorial panel summary, and the representation sensitivity files. It compares
the rebuilt files byte-for-byte with the distributed versions and compiles the
manuscript. In operational terms, this checks that the declared aggregate
and manuscript dependencies are present, rebuilds the aggregate-derived
outputs, and compiles the supplied manuscript source in a clean directory.
The figure is a native-LaTeX vector artifact whose plotted inputs are drawn
from the frozen lattice and its reconciled table; its source-level outcome is
also distinguishable by line weight and marker form rather than color alone.

These checks support aggregate-level inspection and a clean-directory compile
of the manuscript package. They do not constitute a raw-data end-to-end
reproduction, re-certify extraction of any source corpus, or recover the
environment in which the original corpus-wide run was performed. The
environment record therefore describes the aggregate check only and does not
promise cross-platform byte identity.

\subsection{Raw-pipeline and parser boundary}

The raw corpora are not part of the public materials. Nor does the recovered
project contain the complete inherited protocol tree, the original locked
execution environment, a complete inventory of raw inputs, or a one-command
path from those inputs to the paper. Consequently, the release does not
provide a complete raw-data rerun. The aggregate evidence is preserved as a
frozen historical record, while the distinction between that record and a
future raw rerun remains explicit.

One recovered Human Know-How loader also has an unresolved decoding risk. It
applies \texttt{unicode\_escape} while parsing labels, which can introduce
mojibake before text normalization. Affected normalized text could in turn
change content-duplicate edges and the component structure under the
content-only and union configurations. No corrected-parser raw rerun is
available, so neither the direction nor the magnitude of possible drift in
the frozen Human Know-How metrics is known. The current evidence therefore
neither establishes that the parser changed the result nor proves that a
corrected parser would preserve it. This unresolved path is also why the
recovered raw runner is retained only as historical audit code and is not
presented as a trusted public execution entry point.
Section~\ref{sec:limitations} states the corresponding interpretive limit on
the frozen qualifier set.

\subsection{Public contents and data access}

The companion aggregate artifact is publicly available in the HieraRepair
GitHub repository at
\href{https://github.com/joy91269/hierarepair}{\texttt{github.com/joy91269/hierarepair}}.
The public materials consist of two parts. The companion aggregate artifact
provides the frozen aggregate evidence, generated scientific tables and
Figure~\ref{fig:section3-union-collapse}, generation and validation scripts,
and dependency records for aggregate inspection. Its existing integrity
records apply to those frozen records and generated scientific assets. The
final manuscript source package provides the final TeX source, bibliography,
figures and tables, and the two deterministic editorial derivatives with
their generator. It is checked by clean extraction, dependency presence,
compilation, citation resolution, font preflight, and derived-file
byte-for-byte comparison; it is not described as covered by the earlier
aggregate integrity records.

Third-party raw corpora are not redistributed. Their acquisition and use
remain governed by the original sources and their respective license terms;
where the current project record does not verify a license or acquisition
path, that status remains unresolved rather than inferred from repository
availability. Retrieved literature PDFs and extracted full text used during
the bounded prior-art review are likewise excluded. Historical and
superseded review materials are not included in the final source archive.
Public literature records retain only bibliographic metadata, source
identity, locators, and the evidence boundaries needed to audit manuscript
citations.

This separation makes the availability statement testable. A reader can
inspect the aggregate evidence chain and compile the manuscript from its
source package, but must obtain third-party corpora independently for any
future raw-level implementation. Neither public component substitutes for a
complete raw-input inventory, a trusted raw pipeline, or a corrected-parser
rerun; the Human Know-How parser risk therefore remains open.

\enlargethispage{2\baselineskip}
\section{Conclusion}
\label{sec:conclusion}

Prior work has already shown that merging relation types can form a giant
component, obstruct component-disjoint splitting, and motivate deletion or
community-based repair \cite{guvenilir2023_dti}. This study narrows the
question to a symmetric measurement across hierarchical procedural corpora:
what component structure is observed when a content-near-duplicate layer and
a common-container layer are evaluated separately and then combined under
transitive closure?

Under the operational source-level rule, \QualifyingSources{} of
\EligibleSources{} gate-eligible sources qualify. MyFixit and Doc2Dial each
retain comparatively dispersed component structure in the individual layers
but become highly concentrated under their union. The remaining sources serve
as explicit negative cases, including WIQA, for which union formation does not
produce the same collapse pattern. The \QualifyingRatio{} result is a
descriptive fraction of an outcome-enriched, non-random panel; it is neither a
population estimate nor evidence that every relation union behaves similarly.

The mechanism analysis sets a separate boundary. A bridge-specific predictor
is associated with the qualifier pattern, but the registered
\texttt{mean\_union\_degree} control satisfies the same separation and
outcome-association conditions. The panel therefore does not identify a
bridge-specific mechanism. The post hoc representation-sensitivity bounds do
not change that conclusion: they preserve the qualitative negative-control
outcome within this panel while also showing that mean degree itself depends
on the clique representation. Likewise, the discrete $\tau$ sweep gives a
negative result under its multiplicity-adjusted decision rule, not an
equivalence result or evidence of a zero threshold effect.

The accompanying diagnostics remain local to their recorded evidence. The
coverage-exposure quantity approximately re-expresses annotated-field
incompleteness in its two audited corpora because candidate-pair weights are
similar across complete and incomplete units. Structured unresolved status
does not imply absence of every frozen-vocabulary cue, but lexical matches are
not semantic labels. Aggregate records support inspection and recomputation
of the reported quantities; the recovered materials do not provide a complete
raw-data-to-paper rerun, and the Human Know-How decoding risk remains
unresolved.

These boundaries define the paper as a measurement and audit rather than a
new splitting method or a general causal account. Future studies can improve
mechanism discrimination by assessing predictor geometry before outcome
testing and recruiting result-blind sources on which candidate and control
predictors disagree. Within this panel, layer-wise acceptability was not
sufficient evidence of union-level capacity; the union therefore merits audit
as an object in its own right.

\section*{AI assistance disclosure}
OpenAI ChatGPT/Codex and Anthropic Claude were used to assist with code and
evidence auditing, literature organization, and manuscript drafting and
editing. The authors reviewed the resulting analyses, citations, and text and
take full responsibility for the final content.

\bibliographystyle{plain}
\bibliography{refs}

@inproceedings{steshin2023_p4r24,
  author    = {Simon Steshin},
  title     = {{{Lo-Hi}: Practical {ML} Drug Discovery Benchmark}},
  booktitle = {Advances in Neural Information Processing Systems},
  volume    = {36},
  pages     = {64526--64554},
  year      = {2023},
  doi       = {10.52202/075280-2816},
  url       = {https://proceedings.neurips.cc/paper_files/paper/2023/hash/cb82f1f97ad0ca1d92df852a44a3bd73-Abstract-Datasets_and_Benchmarks.html}
}

@article{joeres2025_p4r25,
  author  = {Roman Joeres and David B. Blumenthal and Olga V. Kalinina},
  title   = {Data Splitting to Avoid Information Leakage with {{DataSAIL}}},
  journal = {Nature Communications},
  volume  = {16},
  pages   = {3337},
  year    = {2025},
  doi     = {10.1038/s41467-025-58606-8}
}

@article{teufel2023_graphpart,
  author  = {Felix Teufel and Magn{\'u}s Halld{\'o}r G{\'i}slason and Jos{\'e} Juan Almagro Armenteros and Alexander Rosenberg Johansen and Ole Winther and Henrik Nielsen},
  title   = {{{GraphPart}: homology partitioning for biological sequence analysis}},
  journal = {NAR Genomics and Bioinformatics},
  volume  = {5},
  number  = {4},
  pages   = {lqad088},
  year    = {2023},
  doi     = {10.1093/nargab/lqad088}
}

@article{li2026_lppdbbind,
  author  = {Jie Li and Xingyi Guan and Oufan Zhang and Kunyang Sun and Yingze Wang and Dorian Bagni and Teresa Head-Gordon},
  title   = {Leak Proof {{PDBBind}}: A Reorganized Data Set of Protein--Ligand Complexes for More Generalizable Binding Affinity Prediction},
  journal = {The Journal of Physical Chemistry B},
  volume  = {130},
  number  = {2},
  pages   = {730--740},
  year    = {2026},
  doi     = {10.1021/acs.jpcb.5c08598}
}

@inproceedings{bushuiev2024_ppi,
  author    = {Anton Bushuiev and Roman Bushuiev and Ji{\v{r}}{\'i} Sedl{\'a}{\v{r}} and Tom{\'a}{\v{s}} Pluskal and Ji{\v{r}}{\'i} Damborsk{\'y} and Stanislav Mazurenko and Josef Sivic},
  title     = {Revealing Data Leakage in Protein Interaction Benchmarks},
  booktitle = {GEM Workshop at ICLR},
  year      = {2024},
  url       = {https://openreview.net/forum?id=ORMXYUK5IY}
}

@article{durairaj2024_plinder,
  author  = {Janani Durairaj and Yusuf Adeshina and Zhonglin Cao and Xuejin Zhang and Vladas Oleinikovas and Thomas Duignan and Zachary McClure and Xavier Robin and Gabriel Studer and Daniel Kovtun and Emanuele Rossi and Guoqing Zhou and Srimukh Veccham and Clemens Isert and Yuxing Peng and Prabindh Sundareson and Mehmet Akdel and Gabriele Corso and Hannes St{\"a}rk and Gerardo Tauriello and Zachary Carpenter and Michael Bronstein and Emine Kucukbenli and Torsten Schwede and Luca Naef},
  title   = {{{PLINDER}: The protein--ligand interactions dataset and evaluation resource}},
  journal = {bioRxiv},
  year    = {2024},
  doi     = {10.1101/2024.07.17.603955},
  url     = {https://www.biorxiv.org/content/10.1101/2024.07.17.603955v3}
}

@article{lavertu2026_refnd,
  author  = {Anthony Lavertu and Jacob C{\^o}t{\'e} and Jacques Corbeil and Sophie Gobeil and Pascal Germain},
  title   = {{{Refnd}: Preventing Data Leakage in Relational Datasets}},
  journal = {arXiv preprint arXiv:2607.19376},
  year    = {2026},
  url     = {https://arxiv.org/abs/2607.19376}
}

@techreport{gu2003_p4r06,
  author      = {Lifang Gu and Rohan Baxter and Deanne Vickers and Chris Rainsford},
  title       = {Record Linkage: Current Practice and Future Directions},
  institution = {CSIRO Mathematical and Information Sciences},
  number      = {03/83},
  year        = {2003}
}

@article{baas2021_p4r11,
  author  = {Jurian Baas and Mehdi Dastani and Ad Feelders},
  title   = {Exploiting Transitivity Constraints for Entity Matching in Knowledge Graphs},
  journal = {arXiv preprint arXiv:2104.12589},
  year    = {2021},
  url     = {https://arxiv.org/abs/2104.12589}
}

@inproceedings{froebe2020_nearduplicates,
  author    = {Maik Fr{\"o}be and Jan Philipp Bittner and Martin Potthast and Matthias Hagen},
  title     = {The Effect of Content-Equivalent Near-Duplicates on the Evaluation of Search Engines},
  booktitle = {Advances in Information Retrieval},
  series    = {Lecture Notes in Computer Science},
  volume    = {12036},
  pages     = {12--19},
  year      = {2020},
  doi       = {10.1007/978-3-030-45442-5_2}
}

@inproceedings{buckley2004_bpref,
  author    = {Chris Buckley and Ellen M. Voorhees},
  title     = {Retrieval Evaluation with Incomplete Information},
  booktitle = {Proceedings of the 27th Annual International ACM SIGIR Conference on Research and Development in Information Retrieval},
  pages     = {25--32},
  year      = {2004},
  doi       = {10.1145/1008992.1009000}
}

@article{guvenilir2023_dti,
  author  = {Heval Atas Guvenilir and Tunca Do\u{g}an},
  title   = {How to approach machine learning-based prediction of drug/compound--target interactions},
  journal = {Journal of Cheminformatics},
  volume  = {15},
  pages   = {16},
  year    = {2023},
  doi     = {10.1186/s13321-023-00689-w}
}

@article{newman2001_p4r01,
  author  = {Mark E. J. Newman and Steven H. Strogatz and Duncan J. Watts},
  title   = {Random Graphs with Arbitrary Degree Distributions and Their Applications},
  journal = {Physical Review E},
  volume  = {64},
  number  = {2},
  pages   = {026118},
  year    = {2001},
  doi     = {10.1103/PhysRevE.64.026118},
  url     = {https://arxiv.org/abs/cond-mat/0007235}
}

\end{document}